\documentclass[12pt,a4paper]{article}
\usepackage{amssymb}

\makeatletter
\newcount\@hour\newcount\@minute\chardef\@x10\chardef\@xv60
\def\tcitime{
\def\@time{%
  \@minute\time\@hour\@minute\divide\@hour\@xv
  \ifnum\@hour<\@x 0\fi\the\@hour:%
  \multiply\@hour\@xv\advance\@minute-\@hour
  \ifnum\@minute<\@x 0\fi\the\@minute
  }}%
\def\QCTOpt[#1]#2{%
  \def\QCTOptB{#1}
  \def\QCTOptA{#2}
}
\def\QCTNOpt#1{%
  \def\QCTOptA{#1}
  \let\QCTOptB\empty
}
\def\Qct{%
  \@ifnextchar[{%
    \QCTOpt}{\QCTNOpt}
}
\def\QCBOpt[#1]#2{%
  \def\QCBOptB{#1}
  \def\QCBOptA{#2}
}
\def\QCBNOpt#1{%
  \def\QCBOptA{#1}
  \let\QCBOptB\empty
}
\def\Qcb{%
  \@ifnextchar[{%
    \QCBOpt}{\QCBNOpt}
}
\def\PrepCapArgs{%
  \ifx\QCBOptA\empty
    \ifx\QCTOptA\empty
      {}%
    \else
      \ifx\QCTOptB\empty
        {\QCTOptA}%
      \else
        [\QCTOptB]{\QCTOptA}%
      \fi
    \fi
  \else
    \ifx\QCBOptA\empty
      {}%
    \else
      \ifx\QCBOptB\empty
        {\QCBOptA}%
      \else
        [\QCBOptB]{\QCBOptA}%
      \fi
    \fi
  \fi
}
\newcount\GRAPHICSTYPE
\GRAPHICSTYPE=\z@
\def\GRAPHICSPS#1{%
 \ifcase\GRAPHICSTYPE%\GRAPHICSTYPE=0
   \special{ps: #1}%
 \or%\GRAPHICSTYPE=1
   \special{language "PS", include "#1"}%
 \fi
}%
\def\graffile#1#2#3#4{%
    \leavevmode
    \raise -#4 \BOXTHEFRAME{%
        \hbox to #2{\raise #3\hbox{\null #1}}}%
}%
\def\draftbox#1#2#3#4{%
 \leavevmode\raise -#4 \hbox{%
  \frame{\rlap{\protect\tiny #1}\hbox to #2%
   {\vrule height#3 width\z@ depth\z@\hfil}%
  }%
 }%
}%
\newcount\draft
\draft=\z@

\newif\ifwasdraft
\wasdraftfalse

\def\GRAPHIC#1#2#3#4#5{%
 \ifnum\draft=\@ne\draftbox{#2}{#3}{#4}{#5}%
  \else\graffile{#1}{#3}{#4}{#5}%
  \fi
 }%
\def\addtoLaTeXparams#1{%
    \edef\LaTeXparams{\LaTeXparams #1}}%
\newif\ifBoxFrame \BoxFramefalse
\newif\ifOverFrame \OverFramefalse
\newif\ifUnderFrame \UnderFramefalse

\def\BOXTHEFRAME#1{%
   \hbox{%
      \ifBoxFrame
         \frame{#1}%
      \else
         {#1}%
      \fi
   }%
}

\def\doFRAMEparams#1{\BoxFramefalse\OverFramefalse\UnderFramefalse\readFRAMEparams#1\end}%
\def\readFRAMEparams#1{%
 \ifx#1\end%
  \let\next=\relax
  \else
  \ifx#1i\dispkind=\z@\fi
  \ifx#1d\dispkind=\@ne\fi
  \ifx#1f\dispkind=\tw@\fi
  \ifx#1t\addtoLaTeXparams{t}\fi
  \ifx#1b\addtoLaTeXparams{b}\fi
  \ifx#1p\addtoLaTeXparams{p}\fi
  \ifx#1h\addtoLaTeXparams{h}\fi
  \ifx#1X\BoxFrametrue\fi
  \ifx#1O\OverFrametrue\fi
  \ifx#1U\UnderFrametrue\fi
  \ifx#1w
    \ifnum\draft=1\wasdrafttrue\else\wasdraftfalse\fi
    \draft=\@ne
  \fi
  \let\next=\readFRAMEparams
  \fi
 \next
 }%
\def\IFRAME#1#2#3#4#5#6{%
      \bgroup
      \let\QCTOptA\empty
      \let\QCTOptB\empty
      \let\QCBOptA\empty
      \let\QCBOptB\empty
      #6%
      \parindent=0pt%
      \leftskip=0pt
      \rightskip=0pt
      \setbox0 = \hbox{\QCBOptA}%
      \@tempdima = #1\relax
      \ifOverFrame
          \typeout{This is not implemented yet}%
          \show\HELP
      \else
         \ifdim\wd0>\@tempdima
            \advance\@tempdima by \@tempdima
            \ifdim\wd0 >\@tempdima
               \textwidth=\@tempdima
               \setbox1 =\vbox{%
                  \noindent\hbox to \@tempdima{\hfill\GRAPHIC{#5}{#4}{#1}{#2}{#3}\hfill}\\%
                  \noindent\hbox to \@tempdima{\parbox[b]{\@tempdima}{\QCBOptA}}%
               }%
               \wd1=\@tempdima
            \else
               \textwidth=\wd0
               \setbox1 =\vbox{%
                 \noindent\hbox to \wd0{\hfill\GRAPHIC{#5}{#4}{#1}{#2}{#3}\hfill}\\%
                 \noindent\hbox{\QCBOptA}%
               }%
               \wd1=\wd0
            \fi
         \else
            \ifdim\wd0>0pt
              \hsize=\@tempdima
              \setbox1 =\vbox{%
                \unskip\GRAPHIC{#5}{#4}{#1}{#2}{0pt}%
                \break
                \unskip\hbox to \@tempdima{\hfill \QCBOptA\hfill}%
              }%
              \wd1=\@tempdima
           \else
              \hsize=\@tempdima
              \setbox1 =\vbox{%
                \unskip\GRAPHIC{#5}{#4}{#1}{#2}{0pt}%
              }%
              \wd1=\@tempdima
           \fi
         \fi
         \@tempdimb=\ht1
         \advance\@tempdimb by \dp1
         \advance\@tempdimb by -#2%
         \advance\@tempdimb by #3%
         \leavevmode
         \raise -\@tempdimb \hbox{\box1}%
      \fi
      \egroup%
}%
\def\DFRAME#1#2#3#4#5{%
 \begin{center}
     \let\QCTOptA\empty
     \let\QCTOptB\empty
     \let\QCBOptA\empty
     \let\QCBOptB\empty
     \ifOverFrame 
        #5\QCTOptA\par
     \fi
     \GRAPHIC{#4}{#3}{#1}{#2}{\z@}
     \ifUnderFrame 
        \par #5\QCBOptA
     \fi
 \end{center}%
 }%
\def\FFRAME#1#2#3#4#5#6#7{%
 \begin{figure}[#1]%
  \let\QCTOptA\empty
  \let\QCTOptB\empty
  \let\QCBOptA\empty
  \let\QCBOptB\empty
  \ifOverFrame
    #4
    \ifx\QCTOptA\empty
    \else
      \ifx\QCTOptB\empty
        \caption{\QCTOptA}%
      \else
        \caption[\QCTOptB]{\QCTOptA}%
      \fi
    \fi
    \ifUnderFrame\else
      \label{#5}%
    \fi
  \else
    \UnderFrametrue%
  \fi
  \begin{center}\GRAPHIC{#7}{#6}{#2}{#3}{\z@}\end{center}%
  \ifUnderFrame
    #4
    \ifx\QCBOptA\empty
      \caption{}%
    \else
      \ifx\QCBOptB\empty
        \caption{\QCBOptA}%
      \else
        \caption[\QCBOptB]{\QCBOptA}%
      \fi
    \fi
    \label{#5}%
  \fi
  \end{figure}%
 }%
\newcount\dispkind%
\def\FRAME#1#2#3#4#5#6#7#8{%
 \ifnum\draft=\@ne
   \wasdrafttrue
 \else
   \wasdraftfalse%
 \fi
 \def\LaTeXparams{}%
 \dispkind=\z@
 \def\LaTeXparams{}%
 \doFRAMEparams{#1}%
 \ifnum\dispkind=\z@\IFRAME{#2}{#3}{#4}{#7}{#8}{#5}\else
  \ifnum\dispkind=\@ne\DFRAME{#2}{#3}{#7}{#8}{#5}\else
   \ifnum\dispkind=\tw@
    \edef\@tempa{\noexpand\FFRAME{\LaTeXparams}}%
    \@tempa{#2}{#3}{#5}{#6}{#7}{#8}%
    \fi
   \fi
  \fi
  \ifwasdraft\draft=1\else\draft=0\fi{}%
 }%
\def\TEXUX#1{"texux"}

\long\def\QQQ#1#2{%
     \long\expandafter\def\csname#1\endcsname{#2}}%
\@ifundefined{QTP}{\def\QTP#1{}}{}
\@ifundefined{Qlb}{}{}
\@ifundefined{Qlt}{}{}
\long\def\QQA#1#2{}%
\def\QTR#1#2{{\csname#1\endcsname #2}}%(gp) Is this the best?
\def\EXPAND#1[#2]#3{}%
\def\NOEXPAND#1[#2]#3{}%
\def\LaTeXparent#1{}%
\def\ChildStyles#1{}%
\def\ChildDefaults#1{}%
\def\QTagDef#1#2#3{}%
\def\QQfnmark#1{\footnotemark}

\@ifundefined{INDEX}{\def\INDEX#1#2{}{}}{}%
\@ifundefined{SUBINDEX}{\def\SUBINDEX#1#2#3{}{}{}}{}%
\def\initial#1{\bigbreak{\raggedright\large\bf #1}\kern 2\p@
   \penalty3000}%
\@ifundefined{ZZZ}{}{\makeatletter\input gnuindex.sty\makeatother\makeindex\makeatletter}%
\@ifundefined{abstract}{%
 \def\abstract{%
  \if@twocolumn
   \section*{Abstract (Not appropriate in this style!)}%
   \else \small 
   \begin{center}{\bf Abstract\vspace{-.5em}\vspace{\z@}}\end{center}%
   \quotation 
   \fi
  }%
 }{%
 }%
\@ifundefined{endabstract}{\def\endabstract
  {\if@twocolumn\else\endquotation\fi}}{}%
\@ifundefined{maketitle}{\def\maketitle#1{}}{}%
\@ifundefined{affiliation}{\def\affiliation#1{}}{}%
\@ifundefined{proof}{}{}%
\@ifundefined{endproof}{}{}%
\@ifundefined{newfield}{\def\newfield#1#2{}}{}%
\@ifundefined{chapter}{\def\chapter#1{\par(Chapter head:)#1\par }%
 \newcount\c@chapter}{}%
\@ifundefined{part}{\def\part#1{\par(Part head:)#1\par }}{}%
\@ifundefined{section}{\def\section#1{\par(Section head:)#1\par }}{}%
\@ifundefined{subsection}{\def\subsection#1%
 {\par(Subsection head:)#1\par }}{}%
\@ifundefined{subsubsection}{\def\subsubsection#1%
 {\par(Subsubsection head:)#1\par }}{}%
\@ifundefined{paragraph}{\def\paragraph#1%
 {\par(Subsubsubsection head:)#1\par }}{}%
\@ifundefined{subparagraph}{\def\subparagraph#1%
 {\par(Subsubsubsubsection head:)#1\par }}{}%
\@ifundefined{therefore}{}{}%
\@ifundefined{backepsilon}{}{}%
\@ifundefined{yen}{}{}%
\@ifundefined{registered}{%
   \def\registered{\relax\ifmmode{}\r@gistered
                    \else$\m@th\r@gistered$\fi}%
 \def\r@gistered{^{\ooalign
  {\hfil\raise.07ex\hbox{$\scriptstyle\rm\text{R}$}\hfil\crcr
  \mathhexbox20D}}}}{}%
\@ifundefined{Eth}{}{}%
\@ifundefined{eth}{}{}%
\@ifundefined{Thorn}{}{}%
\@ifundefined{thorn}{}{}%
\@ifundefined{degree}{}{}%
\newdimen\theight
\def\Column{%
 \vadjust{\setbox\z@=\hbox{\scriptsize\quad\quad tcol}%
  \theight=\ht\z@\advance\theight by \dp\z@\advance\theight by \lineskip
  \kern -\theight \vbox to \theight{%
   \rightline{\rlap{\box\z@}}%
   \vss
   }%
  }%
 }%
\def\qed{%
 \ifhmode\unskip\nobreak\fi\ifmmode\ifinner\else\hskip5\p@\fi\fi
 \hbox{\hskip5\p@\vrule width4\p@ height6\p@ depth1.5\p@\hskip\p@}%
 }%
\def\miss{\hbox{\vrule height2\p@ width 2\p@ depth\z@}}%
\def\tcol#1{{\baselineskip=6\p@ \vcenter{#1}} \Column}  %
\def\newfmtname{LaTeX2e}
\def\chkcompat{%
   \if@compatibility
   \else
     \usepackage{latexsym}
   \fi
}

\ifx\fmtname\newfmtname
  \DeclareOldFontCommand{\rm}{\normalfont\rmfamily}{\mathrm}
  \DeclareOldFontCommand{\sf}{\normalfont\sffamily}{\mathsf}
  \DeclareOldFontCommand{\tt}{\normalfont\ttfamily}{\mathtt}
  \DeclareOldFontCommand{\bf}{\normalfont\bfseries}{\mathbf}
  \DeclareOldFontCommand{\it}{\normalfont\itshape}{\mathit}
  \DeclareOldFontCommand{\sl}{\normalfont\slshape}{\@nomath\sl}
  \DeclareOldFontCommand{\sc}{\normalfont\scshape}{\@nomath\sc}
  \chkcompat
\fi

\@ifundefined{theorem}{}{}
\@ifundefined{lemma}{}{}
\@ifundefined{corollary}{}{}
\@ifundefined{conjecture}{}{}
\@ifundefined{proposition}{}{}
\@ifundefined{axiom}{}{}
\@ifundefined{remark}{}{}
\@ifundefined{example}{}{}
\@ifundefined{exercise}{}{}
\@ifundefined{definition}{}{}

\@ifundefined{mathletters}{%
  \newcounter{equationnumber}  
  \def\mathletters{%
     \addtocounter{equation}{1}
     \edef\@currentlabel{\theequation}%
     \setcounter{equationnumber}{\c@equation}
     \setcounter{equation}{0}%
     \edef\theequation{\@currentlabel\noexpand\alph{equation}}%
  }
  
}{}

\@ifundefined{BibTeX}{%
    \def\BibTeX{{\rm B\kern-.05em{\sc i\kern-.025em b}\kern-.08em
                 T\kern-.1667em\lower.7ex\hbox{E}\kern-.125emX}}}{}%
\@ifundefined{AmS}%
    {\def\AmS{{\protect\usefont{OMS}{cmsy}{m}{n}%
                A\kern-.1667em\lower.5ex\hbox{M}\kern-.125emS}}}{}%
\@ifundefined{AmSTeX}{}{}%
\let\DOTSI\relax
\def\RIfM@{\relax\ifmmode}%
\def\FN@{\futurelet\next}%
\newcount\intno@
\def\iint{\DOTSI\intno@\tw@\FN@\ints@}%
\def\iiint{\DOTSI\intno@\thr@@\FN@\ints@}%
\def\iiiint{\DOTSI\intno@4 \FN@\ints@}%
\def\idotsint{\DOTSI\intno@\z@\FN@\ints@}%
\def\ints@{\findlimits@\ints@@}%
\newif\iflimtoken@
\newif\iflimits@
\def\findlimits@{\limtoken@true\ifx\next\limits\limits@true
 \else\ifx\next\nolimits\limits@false\else
 \limtoken@false\ifx\ilimits@\nolimits\limits@false\else
 \ifinner\limits@false\else\limits@true\fi\fi\fi\fi}%
\def\multint@{\int\ifnum\intno@=\z@\intdots@                          %1
 \else\intkern@\fi                                                    %2
 \ifnum\intno@>\tw@\int\intkern@\fi                                   %3
 \ifnum\intno@>\thr@@\int\intkern@\fi                                 %4
 \int}%                                                               %5
\def\multintlimits@{\intop\ifnum\intno@=\z@\intdots@\else\intkern@\fi
 \ifnum\intno@>\tw@\intop\intkern@\fi
 \ifnum\intno@>\thr@@\intop\intkern@\fi\intop}%
\def\intic@{%
    \mathchoice{\hskip.5em}{\hskip.4em}{\hskip.4em}{\hskip.4em}}%
\def\negintic@{\mathchoice
 {\hskip-.5em}{\hskip-.4em}{\hskip-.4em}{\hskip-.4em}}%
\def\ints@@{\iflimtoken@                                              %1
 \def\ints@@@{\iflimits@\negintic@
   \mathop{\intic@\multintlimits@}\limits                             %2
  \else\multint@\nolimits\fi                                          %3
  \eat@}%                                                             %4
 \else                                                                %5
 \def\ints@@@{\iflimits@\negintic@
  \mathop{\intic@\multintlimits@}\limits\else
  \multint@\nolimits\fi}\fi\ints@@@}%
\def\intkern@{\mathchoice{\!\!\!}{\!\!}{\!\!}{\!\!}}%
\def\plaincdots@{\mathinner{\cdotp\cdotp\cdotp}}%
\def\intdots@{\mathchoice{\plaincdots@}%
 {{\cdotp}\mkern1.5mu{\cdotp}\mkern1.5mu{\cdotp}}%
 {{\cdotp}\mkern1mu{\cdotp}\mkern1mu{\cdotp}}%
 {{\cdotp}\mkern1mu{\cdotp}\mkern1mu{\cdotp}}}%
\def\RIfM@{\relax\protect\ifmmode}
\def\text{\RIfM@\expandafter\text@\else\expandafter\mbox\fi}
\let\nfss@text\text
\def\text@#1{\mathchoice
   {\textdef@\displaystyle\f@size{#1}}%
   {\textdef@\textstyle\tf@size{\firstchoice@false #1}}%
   {\textdef@\textstyle\sf@size{\firstchoice@false #1}}%
   {\textdef@\textstyle \ssf@size{\firstchoice@false #1}}%
   \glb@settings}

\def\textdef@#1#2#3{\hbox{{%
                    \everymath{#1}%
                    \let\f@size#2\selectfont
                    #3}}}
\newif\iffirstchoice@
\firstchoice@true
\def\Let@{\relax\iffalse{\fi\let\\=\cr\iffalse}\fi}%
\def\vspace@{\def\vspace##1{\crcr\noalign{\vskip##1\relax}}}%
\def\multilimits@{\bgroup\vspace@\Let@
 \baselineskip\fontdimen10 \scriptfont\tw@
 \advance\baselineskip\fontdimen12 \scriptfont\tw@
 \lineskip\thr@@\fontdimen8 \scriptfont\thr@@
 \lineskiplimit\lineskip
 \vbox\bgroup\ialign\bgroup\hfil$\m@th\scriptstyle{##}$\hfil\crcr}%
\def\Sb{_\multilimits@}%
\def\endSb{\crcr\egroup\egroup\egroup}%
\def\Sp{^\multilimits@}%

\newdimen\ex@
\def\rightarrowfill@#1{$#1\m@th\mathord-\mkern-6mu\cleaders
 \hbox{$#1\mkern-2mu\mathord-\mkern-2mu$}\hfill
 \mkern-6mu\mathord\rightarrow$}%
\def\leftarrowfill@#1{$#1\m@th\mathord\leftarrow\mkern-6mu\cleaders
 \hbox{$#1\mkern-2mu\mathord-\mkern-2mu$}\hfill\mkern-6mu\mathord-$}%
\def\leftrightarrowfill@#1{$#1\m@th\mathord\leftarrow
\mkern-6mu\cleaders
 \hbox{$#1\mkern-2mu\mathord-\mkern-2mu$}\hfill
 \mkern-6mu\mathord\rightarrow$}%
\def\overrightarrow{\mathpalette\overrightarrow@}%
\def\overrightarrow@#1#2{\vbox{\ialign{##\crcr\rightarrowfill@#1\crcr
 \noalign{\kern-\ex@\nointerlineskip}$\m@th\hfil#1#2\hfil$\crcr}}}%

\def\overleftarrow{\mathpalette\overleftarrow@}%
\def\overleftarrow@#1#2{\vbox{\ialign{##\crcr\leftarrowfill@#1\crcr
 \noalign{\kern-\ex@\nointerlineskip}$\m@th\hfil#1#2\hfil$\crcr}}}%
\def\overleftrightarrow{\mathpalette\overleftrightarrow@}%
\def\overleftrightarrow@#1#2{\vbox{\ialign{##\crcr
   \leftrightarrowfill@#1\crcr
 \noalign{\kern-\ex@\nointerlineskip}$\m@th\hfil#1#2\hfil$\crcr}}}%
\def\underrightarrow{\mathpalette\underrightarrow@}%
\def\underrightarrow@#1#2{\vtop{\ialign{##\crcr$\m@th\hfil#1#2\hfil
  $\crcr\noalign{\nointerlineskip}\rightarrowfill@#1\crcr}}}%

\def\underleftarrow{\mathpalette\underleftarrow@}%
\def\underleftarrow@#1#2{\vtop{\ialign{##\crcr$\m@th\hfil#1#2\hfil
  $\crcr\noalign{\nointerlineskip}\leftarrowfill@#1\crcr}}}%
\def\underleftrightarrow{\mathpalette\underleftrightarrow@}%
\def\underleftrightarrow@#1#2{\vtop{\ialign{##\crcr$\m@th
  \hfil#1#2\hfil$\crcr
 \noalign{\nointerlineskip}\leftrightarrowfill@#1\crcr}}}%
\def\qopnamewl@#1{\mathop{\operator@font#1}\nlimits@}
\let\nlimits@\displaylimits
\def\setboxz@h{\setbox\z@\hbox}

\def\varlim@#1#2{\mathop{\vtop{\ialign{##\crcr
 \hfil$#1\m@th\operator@font lim$\hfil\crcr
 \noalign{\nointerlineskip}#2#1\crcr
 \noalign{\nointerlineskip\kern-\ex@}\crcr}}}}

 \def\rightarrowfill@#1{\m@th\setboxz@h{$#1-$}\ht\z@\z@
  $#1\copy\z@\mkern-6mu\cleaders
  \hbox{$#1\mkern-2mu\box\z@\mkern-2mu$}\hfill
  \mkern-6mu\mathord\rightarrow$}
\def\leftarrowfill@#1{\m@th\setboxz@h{$#1-$}\ht\z@\z@
  $#1\mathord\leftarrow\mkern-6mu\cleaders
  \hbox{$#1\mkern-2mu\copy\z@\mkern-2mu$}\hfill
  \mkern-6mu\box\z@$}

\def\projlim{\qopnamewl@{proj\,lim}}
\def\injlim{\qopnamewl@{inj\,lim}}
\def\varinjlim{\mathpalette\varlim@\rightarrowfill@}
\def\varprojlim{\mathpalette\varlim@\leftarrowfill@}
\def\varliminf{\mathpalette\varliminf@{}}
\def\varliminf@#1{\mathop{\underline{\vrule\@depth.2\ex@\@width\z@
   \hbox{$#1\m@th\operator@font lim$}}}}
\def\varlimsup{\mathpalette\varlimsup@{}}
\def\varlimsup@#1{\mathop{\overline
  {\hbox{$#1\m@th\operator@font lim$}}}}

\def\stackunder#1#2{\mathrel{\mathop{#2}\limits_{#1}}}%
\begingroup \catcode `|=0 \catcode `[= 1
\catcode`]=2 \catcode `\{=12 \catcode `\}=12
\catcode`\\=12 
|gdef|@alignverbatim#1\end{align}[#1|end[align]]
|gdef|@salignverbatim#1\end{align*}[#1|end[align*]]

|gdef|@alignatverbatim#1\end{alignat}[#1|end[alignat]]
|gdef|@salignatverbatim#1\end{alignat*}[#1|end[alignat*]]

|gdef|@xalignatverbatim#1\end{xalignat}[#1|end[xalignat]]
|gdef|@sxalignatverbatim#1\end{xalignat*}[#1|end[xalignat*]]

|gdef|@gatherverbatim#1\end{gather}[#1|end[gather]]
|gdef|@sgatherverbatim#1\end{gather*}[#1|end[gather*]]

|gdef|@gatherverbatim#1\end{gather}[#1|end[gather]]
|gdef|@sgatherverbatim#1\end{gather*}[#1|end[gather*]]

|gdef|@multilineverbatim#1\end{multiline}[#1|end[multiline]]
|gdef|@smultilineverbatim#1\end{multiline*}[#1|end[multiline*]]

|gdef|@arraxverbatim#1\end{arrax}[#1|end[arrax]]
|gdef|@sarraxverbatim#1\end{arrax*}[#1|end[arrax*]]

|gdef|@tabulaxverbatim#1\end{tabulax}[#1|end[tabulax]]
|gdef|@stabulaxverbatim#1\end{tabulax*}[#1|end[tabulax*]]

|endgroup

\def\align{\@verbatim \frenchspacing\@vobeyspaces \@alignverbatim
You are using the "align" environment in a style in which it is not defined.}

\@namedef{align*}{\@verbatim\@salignverbatim
You are using the "align*" environment in a style in which it is not defined.}
\expandafter\let\csname endalign*\endcsname =\endtrivlist

\def\alignat{\@verbatim \frenchspacing\@vobeyspaces \@alignatverbatim
You are using the "alignat" environment in a style in which it is not defined.}

\@namedef{alignat*}{\@verbatim\@salignatverbatim
You are using the "alignat*" environment in a style in which it is not defined.}
\expandafter\let\csname endalignat*\endcsname =\endtrivlist

\def\xalignat{\@verbatim \frenchspacing\@vobeyspaces \@xalignatverbatim
You are using the "xalignat" environment in a style in which it is not defined.}

\@namedef{xalignat*}{\@verbatim\@sxalignatverbatim
You are using the "xalignat*" environment in a style in which it is not defined.}
\expandafter\let\csname endxalignat*\endcsname =\endtrivlist

\def\gather{\@verbatim \frenchspacing\@vobeyspaces \@gatherverbatim
You are using the "gather" environment in a style in which it is not defined.}

\@namedef{gather*}{\@verbatim\@sgatherverbatim
You are using the "gather*" environment in a style in which it is not defined.}
\expandafter\let\csname endgather*\endcsname =\endtrivlist

\def\multiline{\@verbatim \frenchspacing\@vobeyspaces \@multilineverbatim
You are using the "multiline" environment in a style in which it is not defined.}

\@namedef{multiline*}{\@verbatim\@smultilineverbatim
You are using the "multiline*" environment in a style in which it is not defined.}
\expandafter\let\csname endmultiline*\endcsname =\endtrivlist

\def\arrax{\@verbatim \frenchspacing\@vobeyspaces \@arraxverbatim
You are using a type of "array" construct that is only allowed in AmS-LaTeX.}

\def\tabulax{\@verbatim \frenchspacing\@vobeyspaces \@tabulaxverbatim
You are using a type of "tabular" construct that is only allowed in AmS-LaTeX.}

\@namedef{arrax*}{\@verbatim\@sarraxverbatim
You are using a type of "array*" construct that is only allowed in AmS-LaTeX.}
\expandafter\let\csname endarrax*\endcsname =\endtrivlist

\@namedef{tabulax*}{\@verbatim\@stabulaxverbatim
You are using a type of "tabular*" construct that is only allowed in AmS-LaTeX.}
\expandafter\let\csname endtabulax*\endcsname =\endtrivlist

\def\@@eqncr{\let\@tempa\relax
    \ifcase\@eqcnt \def\@tempa{& & &}\or \def\@tempa{& &}%
      \else \def\@tempa{&}\fi
     \@tempa
     \if@eqnsw
        \iftag@
           \@taggnum
        \else
           \@eqnnum\stepcounter{equation}%
        \fi
     \fi
     \global\tag@false
     \global\@eqnswtrue
     \global\@eqcnt\z@\cr}

 \def\endequation{%
     \ifmmode\ifinner % FLEQN hack
      \iftag@
        \addtocounter{equation}{-1} % undo the increment made in the begin part
        $\hfil
           \displaywidth\linewidth\@taggnum\egroup \endtrivlist
        \global\tag@false
        \global\@ignoretrue   
      \else
        $\hfil
           \displaywidth\linewidth\@eqnnum\egroup \endtrivlist
        \global\tag@false
        \global\@ignoretrue 
      \fi
     \else   
      \iftag@
        \addtocounter{equation}{-1} % undo the increment made in the begin part
        \eqno \hbox{\@taggnum}
        \global\tag@false%
        $$\global\@ignoretrue
      \else
        \eqno \hbox{\@eqnnum}% $$ BRACE MATCHING HACK
        $$\global\@ignoretrue
      \fi
     \fi\fi
 } 

 \newif\iftag@ \tag@false
 
 \def\tag{\@ifnextchar*{\@tagstar}{\@tag}}
 \def\@tag#1{%
     \global\tag@true
     \global\def\@taggnum{(#1)}}
 \def\@tagstar*#1{%
     \global\tag@true
     \global\def\@taggnum{#1}%  
}

\makeatother

\begin{document}

\begin{center}
$\mathbf{P}$\textbf{-MATRIX DESCRIPTION OF INTERACTION OF TWO CHARGED
HADRONS AND LOW-ENERGY NUCLEAR-COULOMB SCATTERING PARAMETERS}

\textbf{\ }

\vspace{1pt}V. A. Babenko \footnote{{\normalsize
\mbox{Electronic address:
pet@gluk.org}}} and N. M. Petrov

\textit{Bogolyubov Institute for Theoretical Physics,}

\textit{National Academy of Sciences of Ukraine, Kiev, Ukraine}

\vspace{1pt}
\end{center}

\noindent The scattering of two charged strongly interacting particles is
described on the basis of the $P$-matrix approach. In the $P$ matrix, it is
proposed to isolate explicitly the background term corresponding to purely
Coulomb interaction, whereby it becomes possible to improve convergence of
the expansions used and to obtain a correct asymptotic behavior of
observables at high energies. The expressions for the purely Coulomb
background $P$ matrix, its poles and residues, and purely Coulomb
eigenfunctions of the $P$-matrix approach are obtained. The nuclear-Coulomb
low-energy scattering parameters of two charged hadrons are investigated on
the basis of this approach combined with the method of isolating the
background $P$ matrix. Simple explicit expressions for the nuclear-Coulomb
scattering length and effective range in terms of the residual $P$ matrix
are derived. These expressions give a general form of the nuclear-Coulomb
low-energy scattering parameters for models of finite-range strong
interaction. Specific applications of the general expressions derived in
this study are exemplified by considering some exactly solvable models of
strong interaction containing hard core repulsion, and, for these models,
the nuclear-Coulomb low-energy scattering parameters for arbitrary values of
the orbital angular momentum are found explicitly. In particular, the
nuclear-Coulomb scattering length and effective range are obtained
explicitly for the boundary-condition model, the model of a hard-core
delta-shell potential, the Margenau model, and the model of hard-core
square-well potential.

\newpage

\begin{center}
\textbf{1. INTRODUCTION} \vspace{1pt}
\end{center}

\vspace{1pt}

The $P$-matrix approach to describing hadron-hadron interaction was first
proposed by Jaffe and Low \lbrack 1\rbrack\ and was then developed in a
number of studies \lbrack 2--4\rbrack . This approach is a modification of
the well-known Wigner-Eisenbud ${\Re }$-matrix theory \lbrack 5, 6\rbrack .
Within the $P$-matrix approach, the scattering amplitude is expressed in
terms of the logarithmic derivative of the wave function at the surface of
the strong-interaction region --- that is, in terms of the $P$ matrix. Here\
it is assumed that the configuration space of the system is broken down into
two regions, the external region, where the interaction can be described in
terms of a two-particle potential, and the internal region, where strong
interaction is dominant. For the $P$ matrix, a so-called dispersion formula
that appears to be its pole expansion and which establishes its energy
dependence can be derived on the basis of quite general assumptions.
Observables can then be described in terms of a finite number of parameters.

In \lbrack 3, 4, 7\rbrack , a method was proposed for explicitly isolating a
free background part in the $P$ matrix. This method is advantageous in that
it simplifies the implementation of the $P$-matrix approach in specific
applications and extends the region of its applicability. The free $P$
matrix, which corresponds to the absence of interaction, was isolated as the
background part in the aforementioned studies. This is natural in dealing
with the scattering of neutral particles. Here, we propose a generalization
of the isolation method to the case involving charged particles, so that
there is long-range Coulomb interaction in the system along with strong
interaction. It is well known that, in this case, scattering theory requires
a nontrivial modification. We show that the idea of explicitly isolating a
background part in the $P$ matrix as put forth in \lbrack 3, 4, 7\rbrack\
can be implemented for charged particles as well and that, for the
background $P$ matrix, it is advisable in this case to take the purely
Coulomb $P$ matrix --- that is, the logarithmic derivative of the regular
Coulomb wave function at the surface of the interaction region. It turns out
that the isolation of the background Coulomb part offers the same advantages
as in the absence of Coulomb interaction.

\vspace{1pt}As an application of the $P$-matrix approach combined with the
method for isolating the background Coulomb $P$ matrix, we study the
scattering length and the effective range for low-energy nuclear-Coulomb
scattering. These parameters are important physical quantities
characterizing the scattering of charged hadrons and light nuclei at low
energies. We obtain simple explicit expressions for the nuclear-Coulomb
low-energy scattering parameters in terms of the parameters of the residual $%
P$ matrix; these expressions make it possible to analyze and evaluate the
nuclear-Coulomb scattering length and effective range and to find them
directly for finite-range strong-interaction potentials. As a matter of
fact, the expressions that we obtain here determine a general form of the
nuclear-Coulomb low-energy scattering parameters for models of strong
interaction of finite range. In \lbrack 8\rbrack , expressions for the
nuclear-Coulomb low-energy scattering parameters in terms of the $P$-matrix
parameters were obtained without resort to the isolation method. Those
expressions are more cumbersome and less convenient in applications than the
present ones.

\vspace{1pt}Much attention has been given to the nuclear-Coulomb low-energy
scattering parameters (see, for example, \lbrack 9--12\rbrack ) since these
physical quantities play an important role in theoretical and experimental
investigations. In some studies (see \lbrack 13--17\rbrack ), these
parameters were determined explicitly for some specific cases of separable
nuclear potentials (in particular, for the Yamaguchi potential). Here, we
find a general form of the nuclear-Coulomb scattering parameters for a
rather broad class of local strong-interaction models --- namely, for models
of finite range. General expressions obtained for the low-energy parameters
make it possible to determine these quantities explicitly for a number of
exactly solvable strong-interaction models containing hard core repulsion.
We emphasize that it is the use of the simple expressions obtained by the
isolation method that made it possible to simplify significantly the
relevant consideration. It should also be noted that the investigation
presented here was performed for an arbitrary value of the orbital angular
momentum $\ell $.

\newpage

\begin{center}
\textbf{2. DISPERSION FORMULA FOR THE }$P$\textbf{\ MATRIX AND
NUCLEAR-COULOMB OBSERVABLES}
\end{center}

\vspace{1pt}For the elastic scattering of two charged strongly interacting
particles, the radial wave function $\psi _{lk}\left( r\right) $ of relative
motion in a state characterized by a specific value of the orbital angular
momentum $\ell $ is regular at the origin and satisfies the radial
Schr\"{o}dinger equation
\begin{equation}
\left[ \frac{d^{2}}{dr^{2}}+E-\frac{l\left( l+1\right) }{r^{2}}-V\left(
r\right) \right] \psi _{lk}\left( r\right) =0  \tag{1}
\end{equation}
with the potential
\begin{equation}
V\left( r\right) =V_{s}\left( r\right) +V_{c}\left( r\right) \,,  \tag{2}
\end{equation}
\vspace{1pt}where $V_{s}\left( r\right) $ is a short-range strong potential
(by assumption, it has a finite range $R$) and $V_{c}\left( r\right) =2\xi
k/r$ is an ordinary Coulomb potential. Here, $\xi $ is the Coulomb parameter
\begin{equation}
\xi \equiv {\/}\frac{{{\mu e_{1}e_{2}}}}{{{\hbar ^{2}}}k}=\frac{1\text{{\/}}%
}{a{{_{B}}}k}\,,  \tag{3}
\end{equation}
where $a{_{B}}$ is the Bohr radius,
\begin{equation}
a{_{B}}\equiv \frac{{{\hbar ^{2}}}}{{{\mu e_{1}e_{2}}}}{\/}\,,  \tag{4}
\end{equation}
$e_{1}$ and $\,e_{2}$ being the charges of the particles involved in the
scattering process. We shall use the system of units where the reduced
Planck constant and the doubled reduced mass are both equal to unity ($%
\,\hbar =2\mu =1$), so that the energy of the relative motion, $E$, is
expressed in terms of the wave number $k$ as $E=k^{2}$.

At infinity, the radial wave function satisfies the scattering boundary
condition
\begin{equation}
\psi _{lk}\left( r\right) \stackunder{r\rightarrow \infty }{\longrightarrow }%
\overline{\psi }_{lk}\left( r\right) \equiv e^{i\delta _{l}\left( k\right)
}\left[ \cos \nu _{l}\left( k\right) \,F_{l}\left( \xi ,kr\right) +\sin \nu
_{l}\left( k\right) \,G_{l}\left( \xi ,kr\right) \right] \,,  \tag{5}
\end{equation}
\vspace{1pt}where $\overline{\psi }_{lk}\left( r\right) $ is the asymptotic
wave function for the continuous spectrum, while $F_{l}\left( \xi ,kr\right)
$ and $G_{l}\left( \xi ,kr\right) $ are, respectively, the regular and the
irregular Coulomb wave functions [18], whose asymptotic behavior at infinity
is given by
\begin{equation}
F_{l}\left( \xi ,kr\right) \stackunder{r\rightarrow \infty }{\longrightarrow
}\sin \left( kr-\xi \ln 2kr-\frac{l\pi }{2}+\sigma _{l}\left( k\right)
\right) \,,  \tag{6}
\end{equation}
\begin{equation}
G_{l}\left( \xi ,kr\right) \stackunder{r\rightarrow \infty }{\longrightarrow 
}\cos \left( kr-\xi \ln 2kr-\frac{l\pi }{2}+\sigma _{l}\left( k\right)
\right) \,.  \tag{7}
\end{equation}
Here, $\sigma _{l}\left( k\right) \equiv \arg \Gamma \left( l+1+i\xi \right) 
$ is the purely Coulomb phase shift, and the total phase shift $\delta
_{l}\left( k\right) $ has the form 
\begin{equation}
\delta _{l}\left( k\right) =\sigma _{l}\left( k\right) +\nu _{l}\left(
k\right) \,,  \tag{8}
\end{equation}
where $\nu _{l}\left( k\right) $ is the nuclear-Coulomb phase shift.

$\vspace{1pt}$The $P$ matrix $P_{l}(E)$ is defined in terms of the
logarithmic derivative of the radial wave function at the surface of the
strong-interaction region ($\,r=R$), 
\begin{equation}
P_{l}(E)\equiv \frac{R\,\psi _{lk}^{\prime }\left( R\right) }{\psi
_{lk}\left( R\right) }\,.  \tag{9}
\end{equation}
In the internal region $\,r\leqslant R$, we introduce a complete set of
orthonormalized eigenfunctions $u_{ln}\left( r\right) $ that satisfy the Schr%
\"{o}dinger equation (1) and the homogeneous boundary conditions 
\begin{equation}
u_{ln}\left( 0\right) =0\,,\,\,\,u_{ln}\left( R\right) =0  \tag{10}
\end{equation}
\vspace{1pt}at the ends of the interval $\left[ 0,R\right] $. Nontrivial
solutions that obey the conditions in (10) exist only at some energy
eigenvalues $E_{ln}$ that are determined by solving the Sturm-Liouville
problem specified by Eqs. (1) and (10). The orthonormalization conditions
have the form 
\begin{equation}
\int\limits_{0}^{R}\,u_{lm}\,u_{ln}\,dr=\delta _{mn}\,.  \tag{11}
\end{equation}

\vspace{1pt}By expanding the wave function in the internal region in a
series in eigenfunctions $u_{ln}\left( r\right) $, we find that, for the $P$
matrix, there is the dispersion formula \lbrack 1,3,4\rbrack\ 
\begin{equation}
P_{l}(E)=P_{l}(0)+\sum_{n=1}^{\infty }\frac{E}{E_{ln}}\frac{\gamma _{ln}^{2}%
}{E-E_{ln}}\,,  \tag{12}
\end{equation}
where 
\begin{equation}
\gamma _{ln}\equiv \sqrt{R}\,u_{ln}^{\prime }\left( R\right) \,.  \tag{13}
\end{equation}

\vspace{1pt}Relation (12) involves a constant $P_{l}(0)$, the $P$ matrix at
zero energy, and it is the isolation of this constant that ensures
convergence of the remaining series. The dispersion formula (12), which
represents a pole expansion of the $P$ matrix, establishes the general form
of its energy dependence. This dependence is completely determined by the
states of the compound system which are characterized by the energy
eigenvalues $E_{ln}$ and the residues $\gamma _{ln}^{2}$. These quantities
in turn are controlled by the physical properties of the system in the
internal region and are independent of energy $E$.

Let us now establish the relation between the $S$ matrix and the $P$ matrix.
For this, we note that the wave function in the external region, $\overline{%
\psi }_{lk}\left( r\right) $, can be represented in the general form 
\begin{equation}
\overline{\psi }_{lk}\left( r\right) =\frac{i}{2}\left[ H_{l}^{\left(
-\right) }\left( \xi ,kr\right) -S_{l}\left( k\right) H_{l}^{\left( +\right)
}\left( \xi ,kr\right) \right] \,,\;r>R\,,  \tag{14}
\end{equation}
where $H_{l}^{\left( \pm \right) }\left( \xi ,kr\right) $ are the Coulomb
Jost solutions given by 
\begin{equation}
H_{l}^{\left( \pm \right) }\left( \xi ,kr\right) =e^{\mp i\sigma _{l}\left(
k\right) }\left[ G_{l}\left( \xi ,kr\right) \pm iF_{l}\left( \xi ,kr\right)
\right] \,,  \tag{15}
\end{equation}
which represent the diverging and the converging waves distorted by the
Coulomb potential. Accordingly, their asymptotic behavior is given by \ \ 
\begin{equation}
H_{l}^{\left( \pm \right) }\left( \xi ,kr\right) \stackunder{r\rightarrow
\infty }{\longrightarrow }e^{\pm i\left( kr-\xi \ln 2kr-\frac{l\pi }{2}%
\right) }\,.  \tag{16}
\end{equation}
By using the matching conditions at the point $r=R$ and definition (9), we
find that the $S$ matrix can be expressed in terms of the $P$ matrix as\ 
\begin{equation}
S_{l}\left( k\right) =S_{l}^{(h)}\left( k\right) \,\frac{P_{l}^{\left(
-\right) }\left( k\right) -P_{l}\left( k\right) }{P_{l}^{\left( +\right)
}\left( k\right) -P_{l}\left( k\right) }\,,  \tag{17}
\end{equation}
where 
\begin{equation}
S_{l}^{(h)}\left( k\right) \equiv \frac{H_{l}^{\left( -\right) }\left( \xi
,kR\right) }{H_{l}^{\left( +\right) }\left( \xi ,kR\right) }\   \tag{18}
\end{equation}
\ is the $S$ matrix corresponding to the scattering on a hard core of radius 
$R$ in the presence of the Coulomb potential and $P_{l}^{\left( \pm \right)
}\left( k\right) $ are the logarithmic derivatives of the diverging and
converging Coulomb waves at the boundary surface, 
\begin{equation}
P_{l}^{\left( \pm \right) }\left( k\right) \equiv \frac{kR\,H_{l}^{\left(
\pm \right) ^{\prime }}\left( \xi ,kR\right) }{H_{l}^{\left( \pm \right)
}\left( \xi ,kR\right) }\,.  \tag{19}
\end{equation}
Hereafter, a prime denotes differentiation with respect to the variable $%
\rho =kR$. The real and the imaginary part of the function $P_{l}^{\left(
+\right) }\left( k\right) $ are usually denoted by $\triangle _{l}\left(
k\right) $ and $s_{l}\left( k\right) $, respectively; since the relation $%
P_{l}^{\left( +\right) }\left( k\right) =\stackrel{\ast }{P}_{l}^{\left(
-\right) }\left( k\right) $ obviously holds, we can write 
\begin{equation}
P_{l}^{\left( \pm \right) }\left( k\right) =\triangle _{l}\left( k\right)
\pm is_{l}\left( k\right) \,,  \tag{20}
\end{equation}
where the functions $s_{l}\left( k\right) $ and $\triangle _{l}\left(
k\right) $ are expressed in terms of the Coulomb functions as \ [19]\ 
\begin{equation}
s_{l}\left( k\right) =\frac{kR}{F_{l}^{2}\left( \xi ,kR\right)
+G_{l}^{2}\left( \xi ,kR\right) }\,,  \tag{21}
\end{equation}
\begin{equation}
\triangle _{l}\left( k\right) =s_{l}\left( k\right) \left[ F_{l}\left( \xi
,kR\right) F_{l}^{^{\prime }}\left( \xi ,kR\right) +G_{l}\left( \xi
,kR\right) G_{l}^{^{\prime }}\left( \xi ,kR\right) \right] \,.  \tag{22}
\end{equation}
\qquad \qquad

With the aid of Eqs. (17) and (20), it can easily be found that the
nuclear-Coulomb phase shift can be represented as 
\begin{equation}
\nu _{l}\left( k\right) =\zeta _{l}\left( k\right) +\arctan \frac{%
s_{l}\left( k\right) }{P_{l}\left( k\right) -\triangle _{l}\left( k\right) }%
\,,  \tag{23}
\end{equation}
where the phase shift $\zeta _{l}\left( k\right) $ for scattering on a hard
core of radius $R$ in the presence of the Coulomb interaction is given by 
\begin{equation}
\zeta _{l}\left( k\right) \equiv -\arctan \frac{F_{l}\left( \xi ,kR\right) }{%
G_{l}\left( \xi ,kR\right) }\,.  \tag{24}
\end{equation}

\vspace{1pt}Expressions (17) and (23) for the observables reveal a
significant drawback of the $P$-matrix approach based on the dispersion
formula (12) as an approximation of the $P$ matrix: if only a finite number
of terms are retained, the observables in question will have an incorrect
asymptotic behavior at high energies. By way of example, we indicate that,
with increasing energy, the phase shift (23) will then behave as the phase
shift $\zeta _{l}\left( k\right) $ for scattering on a hard core; that is,
it will tend to infinity, 
\begin{equation}
\nu _{l}\left( k\right) \stackunder{k\rightarrow \infty }{\longrightarrow }%
-kR+O\left( 1\right) \,.  \tag{25}
\end{equation}
But in fact, the phase shift must vanish at high energies, at least for
regular potentials.

\newpage

\begin{center}
\textbf{3. PURELY COULOMB }$P$\textbf{\ MATRIX}
\end{center}

\vspace{1pt}An incorrect asymptotic behavior of observables at high energies
can be avoided by isolating the background part in the $P$ matrix. In the
presence of Coulomb interaction, we define the background $P$ matrix as the
purely Coulomb $P$ matrix --- that is, in terms of the logarithmic
derivative of the regular Coulomb wave function, 
\begin{equation}
P_{l}^{\left( c\right) }\left( E\right) \equiv \frac{kR\,F_{l}^{\prime
}\left( \xi ,kR\right) }{F_{l}\left( \xi ,kR\right) }\,.  \tag{26}
\end{equation}
We recall that the regular Coulomb wave function $F_{l}\left( \xi ,\rho
\right) $ is expressed in terms of the confluent hypergeometric function $%
\Phi \left( a,b;z\right) $ as [18]\ 
\begin{equation}
F_{l}\left( \xi ,\rho \right) =C_{l}\left( \xi \right) \,e^{-i\rho }\rho
^{l+1}\,\Phi \left( l+1-i\xi ,2l+2;2i\rho \right) \,,  \tag{27}
\end{equation}
where 
\begin{equation}
C_{l}(\xi )\equiv \frac{2^{l}}{{(2l+1)!}}{\/}\,e^{-\pi \xi /2{\/}%
}\,\left\vert \Gamma \left( l+1+i\xi \right) \right\vert  \tag{28}
\end{equation}
is the Coulomb penetrability factor that is a rather complicated function of
energy and which is introduced in order to ensure the required asymptotic
behavior of the function $F_{l}\left( \xi ,\rho \right) $ at infinity [see
Eq. (6)]. If, however, we consider only the internal region $0\leq r\leq R$,
it is more convenient to introduce a solution that does not involve the
factor $C_{l}(\xi )$ and which possesses simpler properties near the origin.
We define such a solution $\phi _{l}\left( \xi ,\rho \right) $ through the
relation 
\begin{equation}
F_{l}\left( \xi ,\rho \right) ={(2l+1)!!}C_{l}\left( \xi \right) \phi
_{l}\left( \xi ,\rho \right) \,,\,  \tag{29}
\end{equation}
where the factor ${(2l+1)!!}$ was introduced in order that, upon switching
the Coulomb interaction off, the function $\phi _{l}\left( \xi ,\rho \right) 
$ reduce to the spherical Riccati-Bessel function: 
\begin{equation}
\phi _{l}\left( 0,\rho \right) =j_{l}\left( \rho \right) \,.  \tag{30}
\end{equation}

The expression for the function $\phi _{l}\left( \xi ,\rho \right) $ in
terms of a confluent hypergeometric function can easily be found with the
aid of Eqs. (27) and (29). The Coulomb $P$ matrix as expressed in terms of
the solution $\phi _{l}\left( \xi ,\rho \right) $ has the form 
\begin{equation}
P_{l}^{\left( c\right) }\left( E\right) =\frac{kR\,\phi _{l}^{^{\prime
}}\left( \xi ,kR\right) }{\phi _{l}\left( \xi ,kR\right) }\,,  \tag{31}
\end{equation}
which is analogous to (26).

The positions $E_{ln}^{\left( c\right) }\equiv k_{ln}^{\left( c\right) ^{2}}$
of the poles of the Coulomb $P$ matrix --- they depend on the Bohr radius $%
a_{B}$ and on the interaction range $R$ ($E_{ln}^{\left( c\right)
}=E_{ln}^{\left( c\right) }\left( a_{B},R\right) $) --- are defined by the
roots of the denominator of the expression on the right-hand side of (31), 
\begin{equation}
\phi _{l}\left( \frac{1}{a_{B}k_{ln}},k_{ln}R\right) =0\,.  \tag{32}
\end{equation}

The Coulomb eigenfunctions $u_{ln}^{\left( c\right) }\left( r\right) $,
which obey the Schr\"{o}dinger equation (1) with the purely Coulomb
potential and the boundary conditions (10), are given by 
\begin{equation}
u_{ln}^{\left( c\right) }\left( r\right) \,=\frac{\gamma _{ln}^{\left(
c\right) }}{\sqrt{R}k_{ln}^{\left( c\right) }}\frac{\phi _{l}\left( \xi
_{ln},k_{ln}r\right) }{\phi _{l}^{^{\prime }}\left( \xi _{ln},k_{ln}R\right) 
}\,,  \tag{33}
\end{equation}
where $\xi _{ln}\equiv 1/a_{B}k_{ln}$. And the parameters $\gamma
_{ln}^{\left( c\right) }$, which are determined from the normalization
condition for the eigenfunctions, can be found if we use the Green's
theorem, 
\begin{equation}
u_{lk_{1}}\left( R\right) u_{lk_{2}}^{\prime }\left( R\right)
-u_{lk_{1}}^{\prime }\left( R\right) u_{lk_{2}}\left( R\right) =\left(
k_{1}^{2}-k_{2}^{2}\right) \int\limits_{0}^{R}\,u_{lk_{1}\,}u_{lk_{2}}\,dr 
\tag{34}
\end{equation}
for two solutions to Eq. (1) that correspond to two different energy values, 
$k_{1}^{2}$ and $k_{2}^{2}$. By substituting (33) into (34), going over to
the limit $k_{1}\rightarrow k_{2}=k_{ln}$ and taking into account Eqs. (11)
and (13), we find that the expression for the Coulomb residues $\gamma
_{ln}^{\left( c\right) ^{2}}$ can be recast into the form 
\begin{equation}
\gamma _{ln}^{\left( c\right) ^{2}}\,=\frac{2\,E_{ln}^{\left( c\right) }}{%
1-\theta _{l}\left( \xi _{ln},k_{ln}R\right) /a_{B}RE_{ln}^{\left( c\right)
}\phi _{l}^{^{\prime }}\left( \xi _{ln},k_{ln}R\right) \,}\,.  \tag{35}
\end{equation}
Here the function $\theta _{l}\left( \xi ,\rho \right) $ is the derivative
of the function $\phi _{l}\left( \xi ,\rho \right) $ with respect to the
parameter $\xi $,\ 
\begin{equation}
\theta _{l}\left( \xi ,\rho \right) \equiv \frac{\partial \phi _{l}\left(
\xi ,\rho \right) }{\partial \xi }\,,  \tag{36}
\end{equation}
and can be directly expressed in terms of a confluent hypergeometric
function.

Thus, we have completely determined the parameters of the Coulomb $P$ matrix
(its poles and residues) and found the Coulomb eigenfunctions. The
dispersion formula for the Coulomb $P$ matrix has the form (12); that is, 
\begin{equation}
P_{l}^{\left( c\right) }(E)=P_{l}^{\left( c\right) }(0)+\sum_{n=1}^{\infty }%
\frac{E}{E_{ln}^{\left( c\right) }}\frac{\gamma _{ln}^{\left( c\right) ^{2}}%
}{E-E_{ln}}\,,  \tag{37}
\end{equation}
where the $P$ matrix at zero energy, $P_{l}^{\left( c\right) }(0)$, is given
by expression 
\begin{equation}
P_{l}^{\left( c\right) }(0)=\frac{z\,I_{2l}\left( z\right) }{I_{2l+1}\left(
z\right) }-l\,,  \tag{38}
\end{equation}
which can be directly obtained from (31) for $k\rightarrow 0$ by using the
expansion of the regular Coulomb wave function and its derivative in terms
of Bessel functions \lbrack 18, 20, 21\rbrack . In expression (38), $I_{\nu
}\left( z\right) $ are modified Bessel functions and the dimensionless
parameter $z$ is given by 
\begin{equation}
z\equiv 2\sqrt{2R/a_{B}}\,.  \tag{39}
\end{equation}
So far, we have considered the case of Coulomb repulsion ($a_{B}>0$). In the
case of Coulomb attraction ($a_{B}<0$), $P_{l}^{\left( c\right) }(0)$ has
the form\ 
\begin{equation}
P_{l}^{\left( c\right) }(0)=\frac{\zeta \,J_{2l}\left( \zeta \right) }{%
J_{2l+1}\left( \zeta \right) }-l\,,  \tag{40}
\end{equation}
where $J_{\nu }\left( \zeta \right) $ are Bessel functions and $\zeta \equiv
2\sqrt{2R/\left| a_{B}\right| }\,$.

\vspace{1pt}

\begin{center}
\textbf{4. ISOLATING THE PURELY COULOMB BACKGROUND }$P$\textbf{\ MATRIX}
\end{center}

In nuclear-Coulomb $P$ matrix (9), we now isolate explicitly the purely
Coulomb background $P$ matrix (26), following a way that is similar to that
used to isolate explicitly the free background $P$ matrix in the absence of
Coulomb interaction \lbrack 3, 4, 7\rbrack . We represent this
transformation in the form 
\begin{equation}
P_{l}(E)=P_{l}^{\left( c\right) }(E)+\widehat{P}_{l}\left( E\right) \,. 
\tag{41}
\end{equation}
With the aid of Eqs. (12) and (37), it can be shown that, for the residual
nuclear-Coulomb $P$ matrix $\widehat{P}_{l}\left( E\right) $, we have the
expansion 
\begin{equation}
\widehat{P}_{l}(E)=\widehat{P}_{l}(0)+\sum_{n=1}^{\infty }\left[ \frac{E}{%
E_{ln}}\frac{\gamma _{ln}^{2}}{E-E_{ln}}-\frac{E}{E_{ln}^{\left( c\right) }}%
\frac{\gamma _{ln}^{\left( c\right) ^{2}}}{E-E_{ln}^{\left( c\right) }}%
\right] \,.  \tag{42}
\end{equation}

By comparing the expansions (12) and (42) for the functions $P_{l}(E)$ and $%
\widehat{P}_{l}(E)$ we conclude that the isolation of the purely Coulomb
background part in the nuclear-Coulomb $P$ matrix according to (41) amounts
to a partial summation of the series in (12), where one isolates the part
that corresponds to the Coulomb interaction and which is known explicitly.
This naturally improves convergence of the original series, thereby making
it possible to obtain a more accurate representation of observables as
functions of energy. It can be shown that the expansion in (42) converges at
the same rate as a series whose general term is proportional to $1/n^{4}$,
while the expansion in (12) converges as $1/n^{2}$ --- that is, much more
slowly.

By making transformation (41) in Eq. (17) and using the relation 
\begin{equation}
P_{l}^{\left( c\right) }(k)-P_{l}^{\left( \pm \right) }(k)=\frac{kR\,e^{\mp
i\sigma _{l}}}{F_{l}\left( \xi ,kR\right) H_{l}^{\left( \pm \right) }\left(
\xi ,kR\right) }\equiv \frac{1}{c_{l}^{\left( \pm \right) }\left( k\right) }%
\,,  \tag{43}
\end{equation}
which can easily be verified for logarithmic derivatives, we can
straightforwardly express the nuclear-Coulomb $S$ matrix $S_{l}^{\left(
cs\right) }\left( E\right) =e^{2\,i\,\nu _{l}\left( k\right) }$ in terms of
the residual $P$ matrix $\widehat{P}_{l}(E)$\ as\ 
\begin{equation}
S_{l}^{\left( cs\right) }\left( E\right) =\frac{1+c_{l}^{\left( -\right)
}\left( k\right) \,\widehat{P}_{l}(E)}{1+c_{l}^{\left( +\right) }\left(
k\right) \,\widehat{P}_{l}(E)}\,,  \tag{44}
\end{equation}
where the functions $c_{l}^{\left( \pm \right) }\left( k\right) $ are
determined according to (43) and obviously satisfy the relation $%
c_{l}^{\left( +\right) }\left( k\right) =\stackrel{\ast }{c}_{l}^{\left(
-\right) }\left( k\right) $. With the aid of (44), we can easily represent
the nuclear-Coulomb phase shift as 
\begin{equation}
\nu _{l}\left( k\right) =-\arctan \frac{F_{l}^{2}\left( \xi ,kR\right) \,%
\widehat{P}_{l}(E)}{kR+F_{l}\left( \xi ,kR\right) G_{l}\left( \xi ,kR\right)
\,\widehat{P}_{l}(E)}\,.  \tag{45}
\end{equation}

From Eqs. (44) and (45), it is obvious that, if only a finite number of
terms are retained in the pole expansion (42) for $\,\widehat{P}_{l}(E)$,
the $S$ matrix and the phase shift will have a correct asymptotic behavior
at high energies: 
\begin{equation}
S_{l}\left( k\right) \stackunder{k\rightarrow \infty }{\longrightarrow }%
1\,,\;\nu _{l}\left( k\right) \stackunder{k\rightarrow \infty }{%
\longrightarrow }0\,.  \tag{46}
\end{equation}

Thus, an isolation of the purely Coulomb background term in the $P$ matrix
leads to a correct asymptotic behavior of the observables at high energies
if the residual $P$ matrix is approximated by a finite number of pole terms.
We can see that the transformation (41) has a transparent mathematical and
physical substantiation and that its application provides the same
advantages as in the absence of Coulomb interaction.

\newpage

\begin{center}
\textbf{5. EXPRESSIONS FOR THE NUCLEAR-COULOMB LOW-ENERGY SCATTERING
PARAMETERS IN TERMS OF THE RESIDUAL }$P$\textbf{\ MATRIX PARAMETERS}
\end{center}

A great number of studies have been devoted to the problem of generalizing
and modifying effective-range theory in the presence of long-range Coulomb
interaction (see, for example, \lbrack 22--27\rbrack ). As a result, the
Coulomb-modified effective-range function $K_{csl}(E)$ was introduced, and
the nuclear-Coulomb scattering length and effective range were determined
for the case of $S$-wave scattering, as well as for scattering in a state
characterized by an arbitrary value of the orbital angular momentum $\ell $.
In \lbrack 22, 25\rbrack , it was shown that, in the case of an arbitrary
orbital angular momentum, the effective-range expansion in the presence of
Coulomb interaction has the form 
\[
K_{l}(E)\equiv (2l+1)!!^{2}\,C_{l}^{2}(\xi )\,k^{2l+1}\,\left[ \cot \nu
_{l}(k)+{\frac{{2\xi }}{{C_{0}^{2}(\xi )}}}\,h(\xi )\right] = 
\]
\begin{equation}
=-{\frac{{1}}{{a_{l}}}}+{{\frac{1}{2}}\,r_{l}\,k^{2}}+{\ldots }\,,  \tag{47}
\end{equation}
where the function $h(\xi )\,$is expressed in terms of the digamma function $%
\psi (z)\equiv \Gamma ^{\prime }(z)/{\Gamma (z)}$ as 
\begin{equation}
h(\xi )\equiv Re\,\psi (1+i\xi )-\ln |\xi |\,.  \tag{48}
\end{equation}

In the complex plane of energy $E$, the nuclear-Coulomb effective-range
function $\,K_{l}(E)$ is analytic in some domain near the origin \lbrack
26\rbrack ; hence, it can be expanded in the Maclaurin series (47) in powers
of $E$ in the vicinity of the point $E=0$. Thus, a special role of the
function $K_{l}(E)\,$is associated with its analyticity near $E=0$. The
nuclear-Coulomb scattering length $a_{l}\,$and effective range $\,r_{l}$ are
determined in this case in terms of the coefficients in the expansion (47)
of the function $K_{l}(E)\,$. We note that, in a large number of studies,
the nuclear-Coulomb quantities, such as $a_{l}$, $\,r_{l}$, $K_{l}$ and
others are equipped with the additional indices $c$ and $s$, which label
parameters associated with, respectively, Coulomb and short-range
interactions. This was done in order to distinguish these quantities from
their counterparts in the absence of Coulomb interaction, which are labeled
only with the index $s$. Since we do not consider here the case where there
is no Coulomb field, the indices $c$ and $s$ are suppressed on all
nuclear-Coulomb quantities.

For a further analysis, it is reasonable to introduce the dimensionless
inverse scattering length $\gamma _{l}$, and the dimensionless effective
range $\rho _{l}$ as 
\begin{equation}
\gamma _{l}\equiv \frac{l!^{2}\,a_{B}^{2l+1}}{4a_{l}}\,,  \tag{49}
\end{equation}
\begin{equation}
\rho _{l}\equiv 3\,l!^{2}\,a_{B}^{2l-1}\,r_{l}\,.  \tag{50}
\end{equation}
In the particular case of scattering by a hard core of radius $R$ ($\psi
_{lk}\left( R\right) =0\,,$ $P_{l}\left( E\right) =\infty $), the
nuclear-Coulomb low-energy scattering parameters $\gamma _{l}$ and $\rho
_{l} $ can easily be found in the explicit form \lbrack 8\rbrack\ 
\begin{equation}
\gamma _{l}^{h}=\frac{K_{\nu }\left( z\right) }{I_{\nu }\left( z\right) }\,,
\tag{51}
\end{equation}
\begin{equation}
\rho _{l}^{h}=1-\mu _{l}\gamma _{l}^{h}+\frac{2\left( \lambda _{l}-\beta
\right) }{I_{\nu }^{2}\left( z\right) }\,,  \tag{52}
\end{equation}
where the superscript $h$ denotes, as previously, a hard core and $I_{\nu
}\left( z\right) $ and $K_{\nu }\left( z\right) \,$are modified Bessel
functions. The constants $\nu $, $\lambda _{l}$, and $\mu _{l}$ are given by 
\begin{equation}
\nu \equiv 2l+1\,,  \tag{53}
\end{equation}
\begin{equation}
\lambda _{l}\equiv l(l+1)\,,  \tag{54}
\end{equation}
\begin{equation}
\mu _{l}\equiv 4\nu \lambda _{l}\,,  \tag{55}
\end{equation}
while the dimensionless parameter $\beta $ is defined as 
\begin{equation}
\beta \equiv \frac{R}{a_{B}}\,.  \tag{56}
\end{equation}
As before, the parameter $z$ has the form (39). In order to render the
expressions presented below less cumbersome, it is convenient to isolate
explicitly, in the low-energy parameters $\gamma _{l}$ and $\rho _{l}$, the
parts that correspond to scattering on a hard core. Accordingly, we set 
\begin{equation}
\gamma _{l}=\gamma _{l}^{h}+\hat{\gamma}_{l}\,,  \tag{57}
\end{equation}
\begin{equation}
\rho _{l}=\rho _{l}^{h}+\hat{\rho}_{l}\,,  \tag{58}
\end{equation}
defining, in this way, the residual low-energy nuclear-Coulomb scattering
parameters $\hat{\gamma}_{l}$ and $\hat{\rho}_{l}$.

Let us further express the nuclear-Coulomb scattering length and effective
range in terms of the residual $P$ matrix. By substituting (45) into (47),
we find that the nuclear-Coulomb effective-range function as expressed in
terms of the residual $P$ matrix is given by 
\begin{equation}
K_{l}\left( E\right) ={(2l+1)!!^{2}C_{l}^{2}(\xi )\,}k^{2l+1}\,\left[ \frac{%
2\xi }{{C_{0}^{2}(\xi )}}h\left( \xi \right) -\frac{G_{l}\left( \xi ,\rho
\right) }{F_{l}\left( \xi ,\rho \right) }-\frac{\rho }{F_{l}^{2}\left( \xi
,\rho \right) \,\widehat{P}_{l}(E)}\right] \,,  \tag{59}
\end{equation}
where, as before, we use the notation $\rho =kR$. Let us expand the
right-hand side of Eq. (59) in a Maclaurin series in powers of energy $%
E=k^{2}$. It is obvious that, as long as we are interested neither in the
shape parameter nor in higher expansion coefficients, it is sufficient to
retain only the terms that are linear in $E$. We further make use of the
known relation for the Coulomb penetration factor \lbrack 18\rbrack , 
\begin{equation}
{\frac{{C_{l}^{2}(\xi )}}{{C_{0}^{2}(\xi )}}}={\frac{{2^{2l}}}{{(2l+1)!^{2}}}%
}{(1^{2}+\xi ^{2})(2^{2}+\xi ^{2})\ldots (l^{2}+\xi ^{2})\,,}  \tag{60}
\end{equation}
and of the asymptotic expression for the function $h(\xi )$ at low energies
\lbrack 18\rbrack , 
\begin{equation}
h(\xi )\stackunder{\xi \rightarrow \infty }{\simeq }{\frac{1}{{12\xi ^{2}}}}+%
{\frac{1}{{120\xi ^{4}}}}+\ldots \,.  \tag{61}
\end{equation}
The expansions of the Coulomb wave functions in power series in energy $E$
were previously studied by many authors \lbrack 20, 21, 28, 29\rbrack . To
terms that are linear in energy, these expansions for the case of Coulomb
repulsion can be written as 
\begin{equation}
F_{l}(\xi ,\rho )={\frac{{(2l+1)!\,C_{l}(\xi )}}{{(2\xi )^{l+1}}}}{\frac{z}{2%
}}\left\{ I_{\nu }(z)-{\frac{{z^{3}}}{{96{\xi ^{2}}}}}\left[ I_{\nu +1}(z)+{%
\frac{{2l}}{z}}I_{\nu +2}(z)\right] +\ldots \right\} \,,  \tag{62}
\end{equation}
\begin{equation}
G_{l}(\xi ,\rho )\simeq {\frac{{(2l+1)!\,C_{l}(\xi )}}{{(2\xi
)^{l}\,C_{0}^{2}(\xi )}}}z\left\{ K_{\nu }(z)+{\frac{{z^{3}}}{{96{\xi ^{2}}}}%
}\left[ K_{\nu +1}(z)-{\frac{{2l}}{z}}K_{\nu +2}(z)\right] +\ldots \right\}
\,.  \tag{63}
\end{equation}

The dispersion relation for the residual $P$ matrix (42) can be recast into
the form 
\begin{equation}
\widehat{P}_{l}(E)=\stackrel{\infty }{\stackunder{n=1}{\sum }}\left[ \frac{%
\gamma _{ln}^{2}}{E-E_{ln}}-\frac{\gamma _{ln}^{\left( c\right) ^{2}}}{%
E-E_{ln}^{\left( c\right) }}\right] \,.  \tag{64}
\end{equation}
This expansion contains no additional parameters and is completely
determined by the quantities $E_{ln}$ and $\gamma _{ln}^{2}$. It can be
shown that the expansion in (64) converges at the same rate as a series
whose general term is proportional to $1/n^{2}$. The analyticity of the
residual $P$ matrix in a vicinity of the point $E=0\,$\ immediately follows
from (64) if all energy eigenvalues differ from zero. The expansion of the
residual $P$ matrix in a power series in energy $E$ can be written in the
form 
\begin{equation}
\widehat{P}_{l}(E)=\widehat{P}_{l}+\widehat{Q}_{l}\rho ^{2}+\ldots \,, 
\tag{65}
\end{equation}
where 
\begin{equation}
\widehat{P}_{l}\equiv \widehat{P}_{l}(0)\,,  \tag{66}
\end{equation}
\begin{equation}
\widehat{Q}_{l}\equiv \frac{1}{R^{2}}\widehat{P}_{l}^{\prime }(0)  \tag{67}
\end{equation}
are dimensionless expansion coefficients. We can easily express the
quantities $\,\widehat{P}_{l}$ and $\widehat{Q}_{l}$ in terms of the $P$%
-matrix parameters as 
\begin{equation}
\widehat{P}_{l}=\stackrel{\infty }{\stackunder{n=1}{\sum }}\left[ \frac{%
\gamma _{ln}^{\left( c\right) ^{2}}}{E_{ln}^{\left( c\right) }}-\frac{\gamma
_{ln}^{2}}{E_{ln}}\right] \,,  \tag{68}
\end{equation}
\begin{equation}
\widehat{Q}_{l}=\frac{1}{R^{2}}\stackrel{\infty }{\stackunder{n=1}{\sum }}%
\left[ \frac{\gamma _{ln}^{\left( c\right) ^{2}}}{E_{ln}^{\left( c\right)
^{2}}}-\frac{\gamma _{ln}^{2}}{E_{ln}^{2}}\right] \,\,.  \tag{69}
\end{equation}
We also note that, on the basis of Eq. (41), the quantities $\widehat{P}_{l}$
and $\,\widehat{Q}_{l}$ can be determined from the relations 
\begin{equation}
P_{l}=P_{l}^{\left( c\right) }+\widehat{P}_{l}\,,  \tag{70}
\end{equation}
\begin{equation}
Q_{l}=Q_{l}^{\left( c\right) }+\widehat{Q}_{l}\,,  \tag{71}
\end{equation}
where 
\begin{equation}
P_{l}\equiv P_{l}(0)\,,  \tag{72}
\end{equation}
\begin{equation}
Q_{l}\equiv {\frac{1}{R^{2}}}\,P_{l}^{\prime }(0)  \tag{73}
\end{equation}
are parameters in the expansion of the $P$ matrix, 
\begin{equation}
P_{l}(E)=P_{l}+Q_{l}\rho ^{2}+\ldots \,.  \tag{74}
\end{equation}
The parameters $P_{l}^{\left( c\right) }$ and $Q_{l}^{\left( c\right) }$ in
the expansion of the purely Coulomb background $P$ matrix can be found
explicitly from Eqs. (26) and (62). The results are 
\begin{equation}
P_{l}^{\left( c\right) }=l+1+\frac{z}{2}\,\frac{I_{\nu +1}\left( z\right) }{%
I_{\nu }\left( z\right) }\,,  \tag{75}
\end{equation}
\begin{equation}
2\beta \left( 3Q_{l}^{\left( c\right) }+1\right) =l\nu +\left( lz-\frac{\mu
_{l}}{z}\right) \frac{I_{\nu +1}\left( z\right) }{I_{\nu }\left( z\right) }%
+2\left( \beta -\lambda _{l}\right) \,\frac{I_{\nu +1}^{2}\left( z\right) }{%
I_{\nu }^{2}\left( z\right) }\,.  \tag{76}
\end{equation}

Substituting now expressions (60)-(63) and (65) into Eq. (59) and taking
into account Eq. (47), we arrive at explicit expressions for the inverse
scattering length $\hat{\gamma}_{l}$ and the effective range $\hat{\rho}_{l}$%
, also referred to as the dimensionless nuclear-Coulomb residual parameters.
The results are given by 
\begin{equation}
\hat{\gamma}_{l}=\frac{1}{2\,I_{\nu }^{2}\,\widehat{P}_{l}}\,,  \tag{77}
\end{equation}
\begin{equation}
\frac{\hat{\rho}_{l}}{z^{2}\,\hat{\gamma}_{l}}=3\beta \,\frac{\widehat{Q}_{l}%
}{\widehat{P}_{l}}+4\,\frac{\lambda _{l}-\beta }{z}\frac{I_{\nu +1}}{I_{\nu }%
}-l\,,  \tag{78}
\end{equation}
where $I_{\nu }\equiv I_{\nu }(z)$. In the particular case of interaction in
the $S$ state ($\,l=0$), the last formulas are somewhat simplified to become 
\begin{equation}
\hat{\gamma}=\frac{1}{2\,I_{1}^{2}\,\widehat{P}}\,,  \tag{79}
\end{equation}
\begin{equation}
\frac{\hat{\rho}}{z^{2}\,\hat{\gamma}}=3\beta \,\frac{\widehat{Q}}{\widehat{P%
}}-\frac{z}{2}\frac{I_{2}}{I_{1}}\,.  \tag{80}
\end{equation}
In the case of Coulomb attraction ($a_{B}<0$), all the above formulas are
valid upon the substitution of conventional Bessel functions for modified
ones: 
\begin{equation}
I_{\nu }(z)\rightarrow {{\frac{1}{{i^{\nu }}}}J_{\nu }(\zeta )},\,\,K_{\nu
}(z)\rightarrow {{\frac{\pi }{{2i^{\nu }}}}Y_{\nu }(\zeta )},\,\,\zeta =2%
\sqrt{{2R}/{|a_{B}|}}\,.  \tag{81}
\end{equation}

Formulas (77) and (78) yield general explicit expressions for the
nuclear-Coulomb low-energy scattering parameters in terms of the residual $P$
matrix parameters. These expressions make it possible to obtain directly a
general form of the nuclear-Coulomb scattering length and effective range
for models of finite-range strong interaction. \ 

\vspace{1pt}

\begin{center}
\textbf{6. NUCLEAR-COULOMB LOW-ENERGY SCATTERING PARAMETERS FOR EXACTLY
SOLVABLE MODELS CONTAINING HARD CORE REPULSION}
\end{center}

For specific applications of the above general expressions, we will consider
some exactly solvable models of finite-range strong interaction containing
hard core repulsion. For these, we will find explicitly the nuclear-Coulomb
low-energy scattering\ parameters for arbitrary values of the orbital
angular momentum.

\begin{center}
\noindent \textit{6.1. Boundary Condition Model}\textbf{\ }
\end{center}

In the boundary condition model, the interaction in the internal region is
determined by a single energy-independent parameter, the value of the
logarithmic derivative of the wave function at the boundary surface --- that
is, the constant $P_{l}$. It is obvious that the parameter $Q_{l}\,$vanishes
in this case. Thus, we have 
\begin{equation}
P_{l}\left( E\right) =P_{l}\,,  \tag{82}
\end{equation}
\begin{equation}
Q_{l}=0\,,  \tag{83}
\end{equation}
whence it follows that 
\begin{equation}
\widehat{P}_{l}=P_{l}-P_{l}^{\left( c\right) }\,,  \tag{84}
\end{equation}
\begin{equation}
\widehat{Q}_{l}=-Q_{l}^{\left( c\right) }\,.  \tag{85}
\end{equation}
The nuclear-Coulomb low-energy scattering parameters can then be written as 
\begin{equation}
\hat{\gamma}_{l}=\frac{1}{2\,I_{\nu }^{2}\,\left( P_{l}-P_{l}^{\left(
c\right) }\right) }\,,  \tag{86}
\end{equation}
\begin{equation}
\frac{\hat{\rho}_{l}}{z^{2}\,\hat{\gamma}_{l}}=3\beta \,\frac{Q_{l}^{\left(
c\right) }}{\left( P_{l}^{\left( c\right) }-P_{l}\right) }+4\,\frac{\lambda
_{l}-\beta }{z}\frac{I_{\nu +1}}{I_{\nu }}-l\,,  \tag{87}
\end{equation}
where the quantities $P_{l}^{\left( c\right) }$ and $Q_{l}^{\left( c\right)
} $ are given by Eqs. (75) and (76).

\vspace{1pt}

\begin{center}
\noindent \textit{6.2. Hard-Core Delta-Shell Potential}
\end{center}

Let us consider the case where the strong interaction is described by the
delta-shell potential concentrated on the sphere of radius $R$ and
supplemented with a hard core of radius $R_{c}$ less than $R$, 
\begin{equation}
V_{s}\left( r\right) =\left\{ 
\begin{array}{l}
+\infty \,,\,r<R_{c} \\ 
-\frac{\lambda }{R}\,\delta \left( r-R\right) \,,\,r>R_{c}\,
\end{array}
\right. \,.  \tag{88}
\end{equation}
Here, $\lambda $ is the dimensionless interaction constant. In this case,
the wave function in the internal region ($\,r<R$) is a linear combination
of the Coulomb wave functions, 
\begin{equation}
\psi _{lk}\left( r\right) =A_{l}\left( k\right) \,\left[ G_{l}\left( \xi
,kR_{c}\right) F_{l}\left( \xi ,kr\right) -F_{l}\left( \xi ,kR_{c}\right)
G_{l}\left( \xi ,kr\right) \right] ,\,\;R_{c}<r<R  \tag{89}
\end{equation}
and satisfies the zero boundary condition at $r=R_{c}\,$: $\,\psi
_{lk}\left( R_{c}\right) =0\,$. At the boundary surface ($r=R$), the wave
function is continuous, but its derivative undergoes a discontinuity, 
\begin{equation}
\psi _{lk}^{\prime }\left( R+0\right) -\psi _{lk}^{\prime }\left( R-0\right)
=-\frac{\lambda }{R}\,\psi _{lk}\left( R\right) \,.  \tag{90}
\end{equation}
By using formulas (89) and (90), we find that, in the case of a hard-core
delta-shell potential, the $P$ matrix can be represented as 
\begin{equation}
P_{l}(E)=\rho \,\frac{G_{l}\left( \xi ,x\right) F_{l}^{\prime }\left( \xi
,\rho \right) -F_{l}\left( \xi ,x\right) G_{l}^{\prime }\left( \xi ,\rho
\right) }{G_{l}\left( \xi ,x\right) F_{l}\left( \xi ,\rho \right)
-F_{l}\left( \xi ,x\right) G_{l}\left( \xi ,\rho \right) }-\lambda \,, 
\tag{91}
\end{equation}
where $\,x\equiv kR_{c}\,$. With the aid of the definition of the background
Coulomb $P$ matrix (26) and the representation in (41), we obtain the
residual $P$ matrix for the potential (88) in the form 
\begin{equation}
\widehat{P}_{l}(E)=\frac{\rho \,F_{l}\left( \xi ,x\right) \,/\,F_{l}\left(
\xi ,\rho \right) }{G_{l}\left( \xi ,x\right) F_{l}\left( \xi ,\rho \right)
-F_{l}\left( \xi ,x\right) G_{l}\left( \xi ,\rho \right) }-\lambda \,. 
\tag{92}
\end{equation}
By means of the expansion of the Coulomb functions (62) and (63), we derive
the parameters of the residual $P$ matrix, $\widehat{P}_{l}$ and $\,\widehat{%
Q}_{l}$, according to (65). The results are 
\begin{equation}
\widehat{P}_{l}=\frac{1}{2}\,\frac{I_{\nu }\left( y\right) /I_{\nu }\left(
z\right) }{K_{\nu }\left( y\right) \,I_{\nu }\left( z\right) -I_{\nu }\left(
y\right) \,K_{\nu }\left( z\right) }-\lambda \,,  \tag{93}
\end{equation}
\[
24\beta ^{2}\widehat{Q}_{l}=\left( lz^{2}-\mu _{l}\right) \left( \lambda +%
\widehat{P}_{l}\right) +\left( \lambda _{l}-\beta \right) \times 
\]
\begin{equation}
\frac{\left( \lambda _{l}-\alpha \right) /\left( \lambda _{l}-\beta \right)
-2z\,\left[ K_{\nu }\left( y\right) \,I_{\nu +1}\left( z\right) +I_{\nu
}\left( y\right) \,K_{\nu +1}\left( z\right) \right] \,I_{\nu }\left(
y\right) /I_{\nu }\left( z\right) +I_{\nu }^{2}\left( y\right) /I_{\nu
}^{2}\left( z\right) }{\left[ K_{\nu }\left( y\right) \,I_{\nu }\left(
z\right) -I_{\nu }\left( y\right) \,K_{\nu }\left( z\right) \right] ^{2}}\,,
\tag{94}
\end{equation}
where $\alpha \equiv R_{c}/a_{B}\,\ $and $\,y\equiv 2\sqrt{2\alpha }\,$. By
substituting (93) and (94) into (77) and (78), we now find that the
nuclear-Coulomb scattering length and effective range for the hard-core
delta-shell potential can be represented as 
\begin{equation}
\widehat{\gamma }_{l}=\frac{\gamma _{l}^{h}\left( y\right) -\gamma
_{l}^{h}\left( z\right) }{1-2\lambda \,I_{\nu }^{2}\left( z\right) \left[
\gamma _{l}^{h}\left( y\right) -\gamma _{l}^{h}\left( z\right) \right] }\,, 
\tag{95}
\end{equation}
\begin{equation}
\frac{\widehat{\rho }_{l}}{\widehat{\gamma }_{l}^{2}}=\frac{\rho
_{l}^{h}\left( y\right) -\rho _{l}^{h}\left( z\right) }{\left[ \gamma
_{l}^{h}\left( y\right) -\gamma _{l}^{h}\left( z\right) \right] ^{2}}%
+2\lambda zI_{\nu }\left( z\right) \,\left[ lzI_{\nu }\left( z\right)
+4\left( \beta -\lambda _{l}\right) I_{\nu +1}\left( z\right) \right] \,, 
\tag{96}
\end{equation}
where $\gamma _{l}^{h}\left( y\right) $ and $\rho _{l}^{h}\left( y\right) $
are the low-energy nuclear-Coulomb parameters for a hard core of radius $%
R_{c}$ (Eqs. (51) and (52) with substitutions $z\rightarrow y$ and $\beta
\rightarrow \alpha $), while $\gamma _{l}^{h}\left( z\right) \equiv \gamma
_{l}^{h}$ and $\rho _{l}^{h}\left( z\right) \equiv \rho _{l}^{h}$ are the
parameters for the hard core of radius $R$ (Eqs. (51), (52)). In the
limiting case of a delta-shell potential without a core ($R_{c}\rightarrow
0, $ $\,y\rightarrow 0$, and $\gamma _{l}^{h}\left( y\right) \rightarrow
\infty $), expressions (95) and (96) reduce to the known expressions for the
low-energy nuclear-Coulomb parameters for scattering on a delta potential
\lbrack 16, 17\rbrack\ 
\begin{equation}
\widehat{\gamma }_{l}=-\frac{1}{2\lambda \,I_{\nu }^{2}\left( z\right) }\,, 
\tag{97}
\end{equation}
\begin{equation}
\frac{\widehat{\rho }_{l}}{\widehat{\gamma }_{l}^{2}}=2\lambda zI_{\nu
}\left( z\right) \left[ lzI_{\nu }\left( z\right) +4\left( \beta -\lambda
_{l}\right) I_{\nu +1}\left( z\right) \right] \,.  \tag{98}
\end{equation}

\vspace{1pt}

\begin{center}
\noindent \textit{6.3. Margenau Model}
\end{center}

In the Margenau model \lbrack 30\rbrack , strong interaction is simulated by
a square-well potential containing hard core repulsion; in addition Coulomb
interaction is assumed to be absent in the internal region. The latter is
justified by the fact that, in the internal region, the Coulomb interaction
is much weaker than strong interaction. Thus, the total interaction in this
model is described by the potential 
\begin{equation}
V\left( r\right) =\left\{ 
\begin{array}{l}
+\infty \,,\,r<R_{c}\,, \\ 
-V_{0}\,,\,R_{c}<r<R\,, \\ 
{{2\xi k}/{r}}\,,\,r>R\,.
\end{array}
\right.  \tag{99}
\end{equation}
In this case, the wave function in the internal region has the form 
\begin{equation}
\psi _{lk}\left( r\right) =A_{l}\left( k\right) \,\left[ n_{l}\left(
KR_{c}\right) j_{l}\left( Kr\right) -j_{l}\left( KR_{c}\right) n_{l}\left(
Kr\right) \right] ,\;\,R_{c}<r<R\,,  \tag{100}
\end{equation}
where $K\equiv \sqrt{V_{0}+E}$, and $\,j_{l}$ and $n_{l}$ are the spherical
Riccati-Bessel functions. The $P$ matrix can then be written as 
\begin{equation}
P_{l}(E)=\rho \,\frac{j_{l}\left( x\right) n_{l}^{\prime }\left( \rho
\right) -n_{l}\left( x\right) j_{l}^{\prime }\left( \rho \right) }{%
j_{l}\left( x\right) n_{l}\left( \rho \right) -n_{l}\left( x\right)
j_{l}\left( \rho \right) }\,,  \tag{101}
\end{equation}
where $x\equiv KR_{c}$ and $\rho \equiv KR$. From the above, we can easily
determine the $P$-matrix parameters $P_{l}$ and $Q_{l}$. The results are 
\begin{equation}
P_{l}=\rho _{0}\,\frac{j_{l}\left( x_{0}\right) n_{l}^{\prime }\left( \rho
_{0}\right) -n_{l}\left( x_{0}\right) j_{l}^{\prime }\left( \rho _{0}\right) 
}{j_{l}\left( x_{0}\right) n_{l}\left( \rho _{0}\right) -n_{l}\left(
x_{0}\right) j_{l}\left( \rho _{0}\right) }\,,  \tag{102}
\end{equation}
\begin{equation}
2Q_{l}+1=\frac{\lambda _{l}+P_{l}-P_{l}^{2}}{\rho _{0}^{2}}+\frac{b}{\left[
j_{l}\left( x_{0}\right) n_{l}\left( \rho _{0}\right) -n_{l}\left(
x_{0}\right) j_{l}\left( \rho _{0}\right) \right] ^{2}}\,,  \tag{103}
\end{equation}
where $\,b\equiv R_{c}/R\,$, $\,x_{0}\equiv K_{0}R_{c}\,$, $\,\rho
_{0}\equiv K_{0}R\,$, and $\,K_{0}\equiv \sqrt{V_{0}}\,$. By using Eqs. (70)
and (71) and taking into account Eqs. (75) and (76), we can find the
parameters of the residual $P$ matrix --- $\widehat{P}_{l}$ and $\widehat{Q}%
_{l}$; the nuclear-Coulomb low-energy scattering parameters in this case are
given by 
\begin{equation}
\frac{1}{I_{\nu }^{2}\widehat{\gamma }_{l}}=-2\rho _{0}\frac{j_{l}\left(
x_{0}\right) n_{l+1}\left( \rho _{0}\right) -n_{l}\left( x_{0}\right)
j_{l+1}\left( \rho _{0}\right) }{j_{l}\left( x_{0}\right) n_{l}\left( \rho
_{0}\right) -n_{l}\left( x_{0}\right) j_{l}\left( \rho _{0}\right) }-z\frac{%
I_{\nu +1}}{I_{\nu }}\,,  \tag{104}
\end{equation}
\[
\frac{\widehat{\rho }_{l}}{z^{2}I_{\nu }^{2}\widehat{\gamma }_{l}^{2}}%
=-\beta -{l}\nu +\frac{\mu _{l}}{z}\frac{I_{\nu +1}}{I_{\nu }}+2\left( \beta
-\lambda _{l}\right) \frac{I_{\nu +1}^{2}}{I_{\nu }^{2}}+3\beta \frac{%
b-\left[ j_{l}\left( x_{0}\right) n_{l+1}\left( \rho _{0}\right)
-n_{l}\left( x_{0}\right) j_{l+1}\left( \rho _{0}\right) \right] ^{2}}{%
\left[ j_{l}\left( x_{0}\right) n_{l}\left( \rho _{0}\right) -n_{l}\left(
x_{0}\right) j_{l}\left( \rho _{0}\right) \right] ^{2}}+ 
\]
\begin{equation}
+\left[ 3\nu \xi _{0}+2l\rho _{0}+\left( \beta -\lambda _{l}\right) \frac{z}{%
\xi _{0}}\frac{I_{\nu +1}}{I_{\nu }}\right] \frac{j_{l}\left( x_{0}\right)
n_{l+1}\left( \rho _{0}\right) -n_{l}\left( x_{0}\right) j_{l+1}\left( \rho
_{0}\right) }{j_{l}\left( x_{0}\right) n_{l}\left( \rho _{0}\right)
-n_{l}\left( x_{0}\right) j_{l}\left( \rho _{0}\right) }\,,  \tag{105}
\end{equation}
where $\,\xi _{0}\equiv {\frac{{1}}{{a_{B}K_{0}}}}\,$.

\vspace{1pt}

\begin{center}
\textit{6.4. Hard-Core Square-Well Potential}
\end{center}

For the case of a strong-interaction potential in the form of a square well
with a hard core, 
\begin{equation}
V_{s}\left( r\right) =\left\{ 
\begin{array}{l}
+\infty \,,\,r<R_{c}\,, \\ 
-V_{0}\,\theta \left( R-r\right) \,,\,r>R_{c}\,
\end{array}
\right.  \tag{106}
\end{equation}
we confine ourselves to determining the nuclear-Coulomb scattering length.
For the simpler case of a square-well potential without a core, the
nuclear-Coulomb scattering length and the nuclear-Coulomb effective range
were found in \lbrack 8\rbrack . In this case, the wave function in the
internal region has the form 
\begin{equation}
\psi _{lk}\left( r\right) =A_{l}\left( k\right) \,\left[ G_{l}\left( \Xi
,KR_{c}\right) F_{l}\left( \Xi ,Kr\right) -F_{l}\left( \Xi ,KR_{c}\right)
G_{l}\left( \Xi ,Kr\right) \right] ,\,\;R_{c}<r<R\,,  \tag{107}
\end{equation}
where $\,\Xi \equiv \frac{1}{{a_{B}K}}\,$, and the $P$ matrix is given by 
\begin{equation}
P_{l}(E)=\rho \,\frac{F_{l}\left( \Xi ,x\right) G_{l}^{\prime }\left( \Xi
,\rho \right) -G_{l}\left( \Xi ,x\right) F_{l}^{\prime }\left( \Xi ,\rho
\right) }{F_{l}\left( \Xi ,x\right) G_{l}\left( \Xi ,\rho \right)
-G_{l}\left( \Xi ,x\right) F_{l}\left( \Xi ,\rho \right) }\,,  \tag{108}
\end{equation}
where, as in the preceding subsection, $x\equiv KR_{c}$ and $\rho \equiv KR$%
. In accordance with (70) and (75), the parameter $\widehat{P}_{l}$ of the
residual $P$ matrix then assumes the form 
\begin{equation}
\widehat{P}_{l}=\rho _{0}\,\frac{F_{l}\left( \xi _{0},x_{0}\right)
G_{l}^{\prime }\left( \xi _{0},\rho _{0}\right) -G_{l}\left( \xi
_{0},x_{0}\right) F_{l}^{\prime }\left( \xi _{0},\rho _{0}\right) }{%
F_{l}\left( \xi _{0},x_{0}\right) G_{l}\left( \xi _{0},\rho _{0}\right)
-G_{l}\left( \xi _{0},x_{0}\right) F_{l}\left( \xi _{0},\rho _{0}\right) }-%
\frac{z}{2}\frac{I_{\nu +1}\left( z\right) }{I_{\nu }\left( z\right) }%
-\left( l+1\right) \,,  \tag{109}
\end{equation}
while the nuclear-Coulomb scattering length is given by 
\begin{equation}
{\frac{1}{{I_{\nu }^{2}\widehat{\gamma }_{l}}}}=2\rho _{0}{\frac{{\
F_{l}\left( \xi _{0},x_{0}\right) G_{l}^{\prime }\left( \xi _{0},\rho
_{0}\right) -G_{l}\left( \xi _{0},x_{0}\right) F_{l}^{\prime }\left( \xi
_{0},\rho _{0}\right) }}{{F_{l}\left( \xi _{0},x_{0}\right) G_{l}\left( \xi
_{0},\rho _{0}\right) -G_{l}\left( \xi _{0},x_{0}\right) F_{l}\left( \xi
_{0},\rho _{0}\right) }}}-z{\frac{{I_{\nu +1}}}{{I_{\nu }}}}-\left( \nu
+1\right) \,.  \tag{110}
\end{equation}

\vspace{1pt}

\begin{center}
\textbf{7. CONCLUSION}
\end{center}

In summary, an explicit isolation of the purely Coulomb background part in
the $P$ matrix leads to a correct asymptotic behavior of physical
observables at high energies when the residual $P$ matrix is approximated by
a finite number of pole terms. Concurrently, the isolation of the background 
$P$ matrix makes it possible to improve convergence of the remaining
expansions. The transformation in (41) has a transparent mathematical and
physical substantiation, and its application provides the same advantages as
in the absence of Coulomb interaction. In addition, the explicit isolation
of the purely Coulomb background part in the $P$ matrix makes it possible to
obtain the simple general expressions (77) and (78) for the nuclear-Coulomb
low-energy scattering parameters in terms of the residual $P$ matrix. With
the aid of these expressions, we can directly calculate the nuclear-Coulomb
scattering length and effective range for finite-range strong-interaction
potentials. If the Schr\"{o}dinger equation for these potentials admits of
an exact solution in the presence of Coulomb interaction, the
nuclear-Coulomb parameters can be found explicitly. In general, the
nuclear-Coulomb low-energy scattering parameters can be obtained for
arbitrary short-range strong-interaction potentials at any value of the
orbital angular momentum $\ell $. On the basis of the expressions derived in
the present study, we have found explicitly the nuclear-Coulomb scattering
length and effective range for the boundary-condition model, for the model
of a hard-core delta-shell potential, for the Margenau model, and for the
hard-core square-well potential at arbitrary values of the orbital angular
momentum.

\vspace{5pt}

\begin{center}
REFERENCES
\end{center}

\begin{enumerate}
\item  R. L. Jaffe and F. E. Low, Phys. Rev. D \textbf{19}, 2105 (1979).

\item  B. L. G. Bakker and P. J. Mulders, Adv. Nucl. Phys. \textbf{17}, 1
(1986).

\item  V. A. Babenko and N. M. Petrov, Ukr. Fiz. Zh. \textbf{32}, 971 (1987).

\item  V. A. Babenko, N. M. Petrov, and A. G. Sitenko, Can. J. Phys. \textbf{%
70}, 252 (1992).

\item  E. P. Wigner and L. Eisenbud, Phys. Rev. \textbf{72}, 29 (1947).

\item  A. M. Lane and R. G. Thomas, Rev. Mod. Phys. \textbf{30}, 257 (1958).

\item  V. A. Babenko and N. M. Petrov, Yad. Fiz. \textbf{45}, 1619 (1987)
\lbrack Sov. J. Nucl. Phys. \textbf{45}, 1004 (1987)\rbrack .

\item  V. A. Babenko and N. M. Petrov, Yad. Fiz. \textbf{59}, 2154 (1996)
\lbrack Phys. At. Nucl. \textbf{59}, 2074 (1996)\rbrack .

\item  L. D. Landau and Ya. A. Smorodinskij, Zh. Eksp. Teor. Fiz. \textbf{14}%
, 269 (1944).

\item  G. F. Chew and M. L. Goldberger, Phys. Rev. \textbf{75}, 1637 (1949).

\item  J. Schwinger, Phys. Rev. \textbf{78}, 135 (1950).

\item  V. D. Mur, A. E. Kudryavtsev, and V. S. Popov, Yad. Fiz. \textbf{37},
1417 (1983) \lbrack Sov. J. Nucl. Phys. \textbf{37}, 844 (1983)\rbrack .

\item  H. van Haeringen, Nucl. Phys. A \textbf{253}, 355 (1975).

\item  H. van Haeringen, J. Math. Phys. \textbf{18}, 927 (1977).

\item  H. van Haeringen and L. P. Kok, Phys. Lett. A \textbf{82}, 317 (1981).

\item  L. P. Kok, J. W. de Maag, H. H. Brouwer, and H. van Haeringen, Phys.
Rev. C \textbf{26}, 2381 (1982).

\item  J. W. de Maag, L. P. Kok, and H. van Haeringen, J. Math. Phys. 
\textbf{25}, 684 (1984).

\item  Handbook of Mathematical Functions, Ed. by M. Abramowitz and I. A.
Stegun (Dover, New York, 1965).

\item  A. G. Sitenko, Theory of Nuclear Reactions (World Sci., Singapore,
1990).

\item  J. G. Beckerley, Phys. Rev. \textbf{67}, 11 (1945).

\item  J. Humblet, Ann. Phys. (N.Y.) \textbf{155}, 461 (1984).

\item  H. A. Bethe, Phys. Rev. \textbf{76}, 38 (1949).

\item  J. D. Jackson and J. M. Blatt, Rev. Mod. Phys. \textbf{22}, 77 (1950).

\item  T. L. Trueman, Nucl. Phys. \textbf{26}, 57 (1961).

\item  E. Lambert, Helv. Phys. Acta \textbf{42}, 667 (1969).

\item  B. Tromborg and J. Hamilton, Nucl. Phys. B \textbf{76}, 483 (1974).

\item  A. M. Badalyan, L. P. Kok, M. I. Polikarpov, and Yu. A. Simonov,
Phys. Rep. \textbf{82}, 31 (1982).

\item  F. L. Yost, J. A. Wheeler, and G. Breit, Phys. Rev. \textbf{49}, 174
(1936).

\item  F. S. Ham, Q. Appl. Math. \textbf{15}, 31 (1957).

\item  H. Margenau, Phys. Rev. \textbf{59}, 37 (1941).
\end{enumerate}

\end{document}